# Large Coercivity in Nanostructured Rare-earth-free Mn$_x$Ga Films

T.J. Nummy,[1] S.P Bennett,[2] T. Cardinal,[1] and D. Heiman[1]

[1]Department of Physics, Northeastern University, Boston, MA 02115
[2]Department of Mechanical Engineering, Northeastern University, Boston, MA 02115

*Abstract*

The magnetic hysteresis of Mn$_x$Ga films exhibit remarkably large coercive fields as high as $\mu_oH_C$ = 2.5 T when fabricated with nanoscale particles of a suitable size and orientation. This coercivity is an order of magnitude larger than in well-ordered epitaxial film counterparts and bulk materials. The enhanced coercivity is attributed to the combination of large magnetocrystalline anisotropy and ~ 50 nm size nanoparticles. The large coercivity is also replicated in the electrical properties through the anomalous Hall effect. The magnitude of the coercivity approaches that found in rare-earth magnets, making them attractive for rare-earth-free magnet applications.

Rare-earth-based magnets provide the backbone of many products, from computers and mobile phones to electric cars and wind-powered generators.[1] But because of the high cost and limited availability of rare-earth and precious elements, which are expensive to mine and process, there is a growing interest in developing new magnetic materials without these elements.[2-4] This critical need has fostered innovative research aimed at the discovery of novel compounds and nanoscale composites that are free of these elements. One of the key properties of super strong magnets is their large magnetocrystalline anisotropy. With this in mind, it is advantageous to search for rare-earth-free ferromagnets that have large anisotropies, such as MnAl and Mn$_x$Ga. However, in order to take advantage of the large anisotropy and make them suitable for applications they must be synthesized with an appropriate composition and structure at the nanoscale level. We find that when the Heusler compound Mn$_x$Ga is synthesized with the proper nanostructuring, a remarkably high coercive field is produced. These results suggest that Mn$_x$Ga is a good candidate for producing materials with enhanced coercive fields aimed at replacing some rare-earth-based magnets in use today.

Heusler compounds possess a rich variety of fascinating and useful properties.[5] Magnetic shape-memory, thermoelectrics, semi and superconductivity, topological insulators, and half-metallicity are a few examples of the special characteristics unfolding from this appealing niche of materials. The binary ferrimagnet Mn$_x$Ga (x = 2 - 3) is one of the simplest Heusler materials, yet has remarkable magnetic properties. Its magnetism is tunable, where the saturation moment is predicted to be variable by as much as a factor of 4 by varying the stoichiometry from x = 2 to x = 3.[6] The structural and magnetic properties of Mn$_x$Ga have been investigated in bulk materials,[6-9] and thin films grown on various substrates including Si,[10,11] GaAs,[12,13] GaN,[**14**] GaSb,[11] Al$_2$O$_3$,[11] MgO,[15,16] and Cr-MgO[17]. When grown on lattice-matched substrates the resulting epitaxial films exhibit an easy-axis perpendicular to the film plane.[12,16,17] These oriented films have led to measurements of large anisotropy fields extrapolating to $\mu_oH_A$ = 10 T or higher.[16,17] An anisotropy constant of K ~ $10^7$ erg/cm$^3$ was determined from $H_A = 2K/M_S$[18,19] and the saturation magnetization $M_S$. In practice, the coercive field is typically limited to $H_C/H_A$ ~ 0.3, which could lead to a coercive field of at least $\mu_oH_C$ ~ 3 T in this material.

In the present work we show that Mn$_x$Ga films can be synthesized with nanoparticles having ~ 50-100 nm dimensions. This is shown to be an appropriate size for generating remarkably large coercive fields, as large as $\mu_oH_C$ = 2.5 T, which is nearly an order of magnitude larger than for well-ordered epitaxial films and bulk samples. The high coercive fields are attributed to the combination of specific nanostructuring and a large intrinsic magnetocrystalline anisotropy. These results point out a new opportunity for developing rare-earth-free magnetic materials.

Thin films of Mn$_x$Ga in the range x = 2 to 3 were grown by molecular beam epitaxy (MBE) on Si (001) substrates having an amorphous native oxide surface layer. A view of the surface of a 20 nm thick Mn$_x$Ga film is shown in the scanning electron microscope (SEM) image in Fig. 1(a). The particle-like structures are seen to have lateral dimensions on the order of ~ 50-100 nm – a size that is crucial for generating high coercivity. Annealing the films produced some alignment of the crystalline axes, as shown in the reflection high energy electron diffraction (RHEED) images of Figs. 1(b-c). Before annealing, Fig. 1(b) shows a predominant random alignment displayed by the polycrystalline-like ring pattern. The nanocrystal orientation increased when the films were annealed at high temperatures. After annealing at 400 °C, a fraction of the ring intensity coalesced into elongated spots as seen in Fig. 1(c). The annealed particles were found to have a distinct D0$_{22}$ crystal structure[6,12] as determined from θ-2θ x-ray diffraction (XRD) shown in Fig. 1(d). The lattice constants of the D0$_{22}$ structure, illustrated in Fig. 1(e), were found to be *a*=0.389 nm and *c*=0.708 nm, indicating a ~ ½ % contraction in lattice constants compared to bulk values[6]. Magnetization measurements were made in a superconducting quantum interference device (SQUID) magnetometer from Quantum Design. It was found that annealing led to increases in the saturation moment by as much as $10^2$, up to values as high as $M_S$ = 130 emu/cm$^3$.

Figure 2 illustrates the large coercivity in the magnetization that arises from the nanostructuring. This figure compares the room temperature magnetization, M(H), of a 20 nm thick nanostructured Mn$_x$Ga film to that of a highly-ordered epitaxial film grown on GaAs. In contrast to the epitaxial sample, M(H) of the nanostructured film has a wide "s"-shaped major hysteresis (irreversibility) loop. The coercive field for the nanostructured film is exceptionally large, $\mu_oH_C$ = 2.5 T. This is much larger than the coercive field for the epitaxial film, $\mu_oH_C$ = 0.36 T, which is typical of other high-quality epitaxial films[11-13,15] and bulk samples[6]. The magnetism of nanostructured films is robust with respect temperature. The inset of Fig. 2 plots the remanent magnetization, $M_R$(H=0), as a function of temperature, $M_R$(T). The moment remains high at T = 400 K demonstrating a Curie temperature well above room temperature[6,8]. The inset also shows the temperature dependence of the coercive field, $H_C$(T), which appears to have a dependence similar to the remanence. To conclude, we note that the observed 2.5 T coercive field is on the order of those found in some important rare-earth permanent magnets, such as Nd$_2$Fe$_{14}$B where $\mu_o H_C$ ~ 2.6 T[20], and SmCo$_5$ where $\mu_o H_C$ ~ 4 T.[21]

The large coercivity of nanostructured Mn$_x$Ga films is also present in the electronic properties through the anomalous Hall effect (AHE). The Hall effect in magnetic materials contains two main contributions, $\rho_H(H) = R_o H + R_1 M(H,T)$, where $R_o = -1/ne$ is the ordinary Hall effect (OHE) coefficient, $n$ the carrier concentration, $R_1$ the AHE coefficient, and M(H,T) the field and temperature-dependent magnetization. For magnetoconductivity measurements, van der Pauw squares and lithographically fabricated Hall bars with Ti-Au contacts were measured using a 14 T Cryogenics Limited Cryo-free magnet. Despite the particle-like growth, all of the films had metallic conductivity with longitudinal resistivities in the range $\rho$ = 200 to 300 $\mu\Omega\cdot$cm, very similar to that measured in epitaxial films.[11,15] Figure 3 shows results of room temperature Hall effect measurements of a 20 nm thick nanostructured Mn$_x$Ga film. In Fig. 3(a) the raw Hall resistivity is plotted as a function of H applied perpendicular to the film. The AHE gives rise to the irreversible open-loop hysteresis, which closes up at $\mu_o H$ = 5 T. At higher fields the resistivity is reversible and linear in field. The dashed line in Fig. 3(a) represents the total reversible component of the resistivity, which may contain some residual nonsaturating moment. From the slope of the reversible component we obtain a lower limit for the effective carrier concentration, $n \geq 8 \times 10^{21}$ cm$^{-3}$, which corresponds to one electron per unit cell. In Figure 3(b) we compare the hysteresis in the AHE with the magnetization. There is excellent agreement between data from the two measurements, including the magnitude of the coercive field for that sample. The AHE has been observed previously in Mn$_x$Ga, but in those studies the magnitude of the coercive field was limited to < 1 T.[11,12,15] At high fields we find that the AHE resistivity reaches a saturation value of $\rho_{AHE}$ = 1.5 $\mu\Omega\cdot$cm, which is a factor of 2 larger than observed for the precious-metal perpendicular moment material $L1_0$-FePt (0.88 $\mu\Omega\cdot$cm).[22]

The dimensions of the nanoscale particles are important for achieving large coercive fields, as particles that are too large support multidomains, while particle that are too small become superparamagnetic. Several mechanisms hinder the field-induced magnetic alignment that leads to hysteresis. In relatively large particles the hysteresis arises from magnetic domains, which are characterized by pinning of domain walls and nucleation of reversed domains. On the other hand, when particles are too small to support multiple domains, the hysteresis arises from coherent reversal of single magnetic domains that are hindered by anisotropy. Single-domain particles must be small enough to prohibit the formation of domain walls. Energy minimization requires that single domain particles must be smaller than several times ($\geq$ 3) the domain wall width. The width of our domain walls is $\delta_w$ ~ 20-30 nm, using $\delta_w$ ~ $2\pi (A/K_1)^{1/2}$ and A ~ 2 x 10$^{-6}$ erg/cm as the exchange stiffness obtained from the magnetization and Curie temperature[19]. A sizeable fraction of the Mn$_x$Ga particles are single domain since the domain wall width is a significant fraction of the particle size. For a single domain particle with uniaxial anisotropy the coercive field depends on the relative orientation between the unique (easy) magnetic axis and the applied field. For a field applied along the unique axis ($\varphi$ = 0) the coercive field is maximum and is equal to the anisotropy field, $H_C/H_A$ = 1, and the remanent magnetization is equal to the saturation magnetization, $M_R/M_S$ = 1, shown by the dashed curve in Fig. 3(c). As the angle between the applied field and the unique axis increases, the coercive field and remanent magnetization collapse, both reaching zero for a field applied at right angles to the unique axis. The Stoner-Wohlfarth (SW) model[23] treats small noninteracting single-domain particles possessing uniaxial anisotropy. Although it models particle shapes consisting of ellipsoids of revolution having an easy axis along the semi-major axis of the ellipsoid, it is nevertheless useful for qualitative comparisons to real systems of particles. Results of the SW model predict that for a completely random distribution of easy axis directions of prolate spheroids (oblate spheroids have $H_C$ = 0), the remanent magnetization is one-half the saturation value, $M_R/M_S$ = 0.50, and the coercive field is approximately one-half the anisotropy field, $H_C/H_A$ = 0.48, shown by the solid curve in Fig. 3(c). However, the hysteresis data of the nanostructured film in Fig. 2 shows a larger remanence ratio of $M_R/M_S$ = 0.78. (Note that this measured ratio is an upper limit as the saturation moment may be larger due to incomplete saturation at the accessible fields.) The larger observed $M_R/M_S$ ratio may be assigned to partial nonrandom alignment of the easy axes, which is confirmed by the spotty RHEED images. This remanence value is equal to that of a SW particle with its unique axis oriented at $\varphi$ = 40 deg to the applied field, where $M_R/M_S$ = 0.77 and $H_C/H_A$ = 0.50, and shown by the dotted curve in Fig. 3(c). Using an upper limit value of $H_C/H_A$ = 0.5 and our observed value of $\mu_o H_C$ = 2.5 T, we estimate a lower limit for the anisotropy field of $\mu_o H_A \geq 5$ T for the nanostructured film, consistent with extrapolated measurements[16,17].

As a final point we consider other contributions to the coercive field. The uniaxial anisotropy is not limited to *magnetocrystalline* anisotropy, $K_1$, arising from spin-orbit interactions, but can also have contributions from *shape*, *stress*, and *surface* uniaxial magnetic anisotropies. We estimate that the shape contribution[18,19] to the coercive field is negligible (~ 0.05 T) and the surface contribution[19] is small ($\leq$ 0.4 T). For the strain contribution, we applied the Williamson-Hall model of XRD line broadening to the data in Fig. 1(d)[24,25] to obtain a value of RMS strain[26] $\varepsilon$ = 0.5 $\pm$ 0.1 %. Although the magnitude of the magnetostriction is not known, the strain could add noticeably to the coercive field (~ $10^0$ T). In conclusion, the dominant anisotropy leading to the high coercive fields appears to be the magnetocrystalline anisotropy, but strain and surface contributions could play a smaller role.

In summary, we show that when $Mn_xGa$ is synthesized with an appropriate nanoscale structure the coercive field is increased by nearly an order of magnitude over that found in well-ordered epitaxial films and bulk samples. Coercive fields as large as $\mu_o H_C$ = 2.5 T were obtained for films grown on Si substrates. The nanostructured films have strained particle-like features ~ 50-100 nm in size. The remarkably large coercivity is attributed to the combination of large intrinsic magnetocrystalline anisotropy and suitable nanostructuring. In addition, the large coercivity is also present in the electrical conductivity through the anomalous Hall effect. These magnetic and magnetotransport properties could find applications in mechanical and electrical devices that require high coercive fields, and understanding their structures could provide a pathway for developing new rare-earth-free magnetic materials.

This work was supported by the National Science Foundation grant DMR-0907007. We thank L. Lewis, Y. Chen and W. Nowak for useful discussions and W. Fowle for assistance with the electron microscopy studies. Added note: after this work was submitted, an article was published describing a coercive field of 2.05 T in $Mn_{1.5}Ga$.[27]

FIG. 1. (Color online) Images and x-ray diffraction of a nanostructured $Mn_xGa$, x = 2.7, film grown on Si substrate. (a) Nanostructures of island-like particles are pictured in the scanning electron microscope (SEM) image of the surface, where the bar is 200 nm. Reflection high energy electron diffraction (RHEED) images: (b) as-grown at 90 °C; and (c) annealed at 400 °C for 60 minutes showing distinct quasi-alignment of crystallites. (d) X-ray θ-2θ diffraction intensity showing $Mn_xGa(112)$, $Mn_xGa(200)$ and $Mn_xGa(224)$ diffraction peaks of the $D0_{22}$ structure. (e) Illustration of crystal structure of $Mn_3Ga$, where Ga atoms are the small grey spheres at the corners and the center.[6]

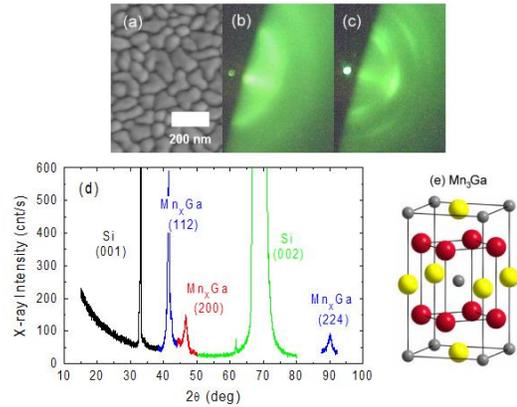

FIG. 2. (Color online) Magnetic properties of a nanostructured $Mn_xGa$, x = 2.7, film illustrating the large coercivity at room temperature. Magnetization (scaled and diamagnetism subtracted) of a nanostructured film grown on Si (blue solid circles) compared to that of a highly-ordered film grown epitaxially on GaAs (red open squares). The solid curves are guides for the eye. The nanostructured film has an order of magnitude larger coercive field, $\mu_oH_C$ = 2.5 T. The inset shows the temperature dependence of the remanent magnetization, $M_R(H=0)$, of the nanostructured film, $M_R(T) / M_R(0)$, (black crosses), and the temperature dependence of the coercive field, $H_C(T) / H_C(0)$, (blue open circles). The solid curves are empirical fits to $(1 - T/T')^\alpha$, where T' = 710 K and α = 0.45 for $M_R(T)$, while T' = 530 K and α = 0.40 for $H_C(T)$.

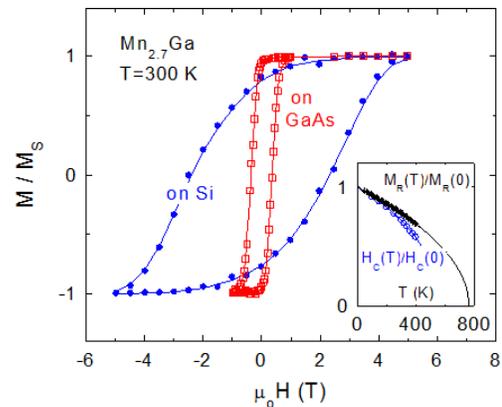

FIG. 3. (Color online) Large coercivity shown in the anomalous Hall effect (AHE) at room temperature for a nanostructured $Mn_xGa$, x = 2.7, film. (a) Solid curve (black) shows the total Hall resistivity, arising from the anomalous Hall effect (AHE) plus reversible resistivity, and the dashed line (red) illustrates the reversible resistivity component. (b) The solid curve shows the AHE resistivity component, and the open (blue) circles show the closely matched magnetization. The magnitudes are scaled by the saturation values to provide comparison. (c) Magnetization of Stoner-Wohlfarth particles for easy-axis aligned with field φ = 0 (black dashed curve), a φ = 40 deg angle between easy axis and field (red dotted curve), and random alignment of easy axes (blue solid curve).

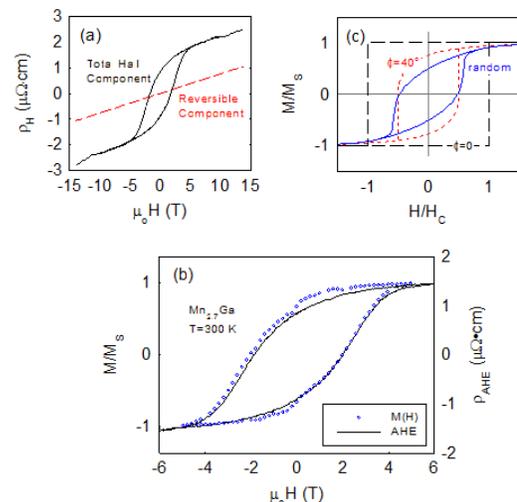

7